\documentclass[conference]{IEEEtran}
\IEEEoverridecommandlockouts
\usepackage{cite}
\usepackage{amsmath,amssymb,amsfonts}
\usepackage{algorithmic}
\usepackage{graphicx}
\usepackage{textcomp}
\usepackage{xcolor}
\usepackage{booktabs}
\usepackage{multirow}
\usepackage{array}
\usepackage{url}
\usepackage{tikz}
\usetikzlibrary{shapes.geometric}
\usepackage{balance}
\usepackage[hidelinks]{hyperref}
\usetikzlibrary{arrows.meta,positioning,calc}

\usepackage{pgfplots}
\pgfplotsset{compat=1.18}
\usetikzlibrary{arrows.meta, positioning, calc}
\usepgfplotslibrary{groupplots, statistics}

\def\BibTeX{{\rm B\kern-.05em{\sc i\kern-.025em b}\kern-.08em
    T\kern-.1667em\lower.7ex\hbox{E}\kern-.125emX}}
\begin{document}

%\title{From Network Intrusion Detection to Blockchain-Backed Endpoint Detection and Response: A Systematization of Knowledge on Decentralized Detection-and-Response Architectures

%\title{The Blockchain-EDR Gap: Rethinking Decentralized Trust Across Intrusion Detection, Prevention, and Endpoint Response

\title{From Network Intrusion Detection to Blockchain-Backed Endpoint Detection and Response: Mapping the Landscape of Decentralized Detection-and-Response Architectures
}

\author{\IEEEauthorblockN{Yahya Shahsavari}
\IEEEauthorblockA{\textit{Dept. of Software and IT Engineering} \\
\textit{École de technologie supérieure (ÉTS)}\\
Montréal, Québec, Canada \\
yahya.shahsavari@etsmtl.ca}
\and
\IEEEauthorblockN{Sara Rouhani}
\IEEEauthorblockA{\textit{Dept. of Computer Science} \\
\textit{University of Calgary}\\
Calgary, Alberta, Canada \\
sara.rouhani@ucalgary.ca}

\and
\IEEEauthorblockN{Kaiwen Zhang}
\IEEEauthorblockA{\textit{Dept. of Software and IT Engineering} \\
\textit{École de technologie supérieure (ÉTS)}\\
Montréal, Québec, Canada \\
kaiwen.zhang@etsmtl.ca}
}

\maketitle

\begin{abstract}
While the literature on blockchain-assisted intrusion detection and prevention systems (IDS/IPS) for Internet of Things (IoT) and Industrial Internet of Things (IIoT) networks is mature, existing systematic reviews suffer from two critical limitations: they overlook the structural shift toward modern Endpoint Detection and Response (EDR) and Extended Detection and Response (XDR) architectures, and they conflate blockchain's distinct functional roles into a single monolithic category. This Systematization of Knowledge (SoK) addresses these gaps by proposing a three-axis taxonomy that classifies proposals by detection-system class (NIDS, HIDS, EDR/XDR), blockchain functional role, and 
response-automation maturity. Synthesizing research published in high-impact venues between 2019 and 2026, we provide a rigorous gap analysis exposing why a genuine per-endpoint blockchain-anchored response loop remains nearly nonexistent due to latency, deployment, and community mismatches. Furthermore, we evaluate structural, cross-cutting challenges persisting across the literature, including consensus latency on constrained devices, post-quantum cryptographic vulnerability, smart-contract attack surfaces, and the adversarial vulnerability of evolving LLM-based detection engines. Finally, we outline a comprehensive research agenda centered on hybrid on-chain/off-chain orchestration to bridge the gap between decentralized trust and rapid response automation.

\end{abstract}


\section{Introduction}\label{sec:intro}

Blockchain technology has attracted sustained interest as a mechanism for enhancing the trustworthiness, resilience, and collaborative capabilities of cyber defense systems. Over the past decade, its integration with intrusion detection and prevention technologies has generated a substantial and steadily expanding body of research. Numerous surveys have systematically reviewed blockchain-enabled intrusion detection systems (IDS) and intrusion prevention systems (IPS), particularly in the context of IoT environments (e.g.~\cite{hassan2023survey,shalabi2024blockchain}). Broader systematic reviews have also synthesized more than one hundred studies spanning blockchain-based IDS, IPS, and intrusion detection and response systems (IDRS) published over multiple years. In parallel, research efforts on collaborative cybersecurity have cataloged blockchain-enabled collaborative IDS architectures across diverse application domains, including healthcare~\cite{asante2025blockchain,chidambaranathan2024deep}, software-defined networking (SDN)~\cite{li2020framework,li2023blockchain,almuseelem2026collaborative}, industrial IoT~\cite{mathew2022integration,nishad2026lightweight,kumar2022p2tif}, smart-city infrastructures~\cite{liu2021artificial,khalil2022blockchain}, and vehicular networks and transportation systems~\cite {peelam2025blockchain,abdel2021federated,shahsavari2026sok}. Among these, the seminal survey by Meng \textit{et al.} remains a widely cited reference for conceptualizing how blockchain's decentralized trust model can be incorporated into IDS architectures~\cite{meng2018intrusion}.

The existence of this extensive literature naturally raises the question of whether another systematization is warranted. Before undertaking this work, we explicitly revisited this question. Our conclusion is affirmative, but for reasons that differ fundamentally from simply incorporating newly published studies. Rather than providing another chronological update of blockchain-based IDS research, this SoK re-examines the field through a broader systems perspective that encompasses modern detection and response technologies and identifies structural shifts that are not adequately captured by existing surveys. Three key observations motivate this perspective.

\textit{Observation 1:} Existing literature is heavily concentrated on IDS/IPS, whereas blockchain-enabled Endpoint EDR remains largely unexplored. Current research on blockchain-enhanced cyber defense is overwhelmingly centered on network-based intrusion detection and IoT-oriented IDS/IPS architectures. The reviewed literature on blockchain-augmented detection is overwhelmingly anchored to network-level and IoT-device-level intrusion detection. Dedicated reviews of EDR technology exist independently~\cite{kaur2024evolution}, and vendor-comparison studies of commercial EDR platforms such as CrowdStrike and SentinelOne have been published as recently as 2026~\cite{eigner2026comparison}, but these EDR-specific reviews treat blockchain, where mentioned at all, as a peripheral data-security or supply-chain-integrity consideration rather than as a structural design choice for the detection-and-response pipeline itself. 
The asymmetry between a mature IoT-IDS-blockchain literature and a nearly nonexistent EDR-blockchain literature is itself a citable, systematizable finding rather than a mere gap statement. Consequently, while blockchain-assisted IDS and IPS constitute a mature and well-surveyed research area, comparable investigations for blockchain-enabled EDR are remarkably scarce. We argue that this imbalance is itself an important finding, reflecting the historical evolution of blockchain security research and motivating a unified examination of blockchain across contemporary detection and response paradigms rather than within IDS alone.

\textit{Observation 2:} "Blockchain-based IDS" conflates distinct functional roles. A closer examination of the literature reveals that the term \emph{blockchain-based IDS} encompasses systems that employ blockchain for fundamentally different purposes. Across existing proposals, blockchain commonly serves at least four distinct architectural roles: (i) immutable storage of alerts, logs, and detection metadata to support auditability and digital forensics~\cite{aliyu2025blockchain}; (ii) decentralized trust establishment and consensus among collaborating detection entities operating without a trusted central authority~\cite{li2021surveying,kolokotronis2019blockchain}; (iii) incentive mechanisms that encourage honest participation and discourage malicious behavior in collaborative detection frameworks through token or reputation-based rewards~\cite{fan2022blockchain,selimi2023cybernfts}; and (iv) integrity assurance for machine learning artifacts, such as verifying model updates exchanged in federated-learning-based intrusion detection systems~\cite{begum2024bflids,ali2025federated,al2022blockchain}. Existing surveys typically organize the literature according to application domains (e.g., IoT, industrial IoT, healthcare,  transportation system, or SDN) or blockchain platforms, rather than according to the architectural function that blockchain performs. As a result, conceptually different systems are frequently grouped together under the same label, obscuring meaningful comparisons and making it difficult to identify which architectural roles have matured and which remain comparatively underexplored.

\textit{Observation 3:} Detection dominates; prevention and response remain shallow. Although the literature is commonly framed as encompassing intrusion detection and prevention, the overwhelming majority of blockchain-enabled security systems focus exclusively on the detection stage. Surveys of collaborative blockchain-based security architectures consistently report that most proposed systems terminate after generating alerts, while comparatively few incorporate explicit intrusion prevention or automated response capabilities~\cite{miller2024collaborative}. Yet many of the characteristics that motivate blockchain adoption (including tamper-resistant state management, decentralized policy enforcement, programmable smart contracts, and verifiable execution) are particularly well suited to supporting coordinated response actions. Examples include automatically quarantining compromised endpoints, enforcing distributed access-control policies, orchestrating containment workflows, and maintaining immutable records of response decisions for auditing and regulatory compliance~\cite{alevizos2025automated}. The limited adoption of these capabilities suggests that current research has largely exploited blockchain as a mechanism for trustworthy detection rather than as an enabling substrate for end-to-end cyber defense. This architectural imbalance motivates a broader examination that spans the entire detection, prevention, and response lifecycle rather than treating intrusion detection in isolation.

These three observations motivate the contribution of this paper:

\begin{itemize}
\item A three-axis taxonomy (Section~\ref{sec:taxonomy}) that cross-classifies the literature by detection-system class, blockchain functional role, and response-automation maturity, making the asymmetries above explicit and quantifiable rather than anecdotal.
\item A structured synthesis (Sections~\ref{sec:ids-ips}--\ref{sec:edr}) of blockchain-IDS/IPS and blockchain-EDR work published in high-impact venues (IEEE Internet of Things Journal, IEEE Transactions on Vehicular Technology, IEEE Transactions on Knowledge and Data Engineering, IEEE Communications Surveys \& Tutorials, ACM Computing Surveys, Computers \& Security, and others) from 2018 through 2026.
\item A cross-cutting analysis (Section~\ref{sec:crosscutting}) of recurring open problems (i.e., scalability/consensus latency, post-quantum cryptographic exposure, and evaluation-methodology weaknesses) that persist across nearly every sub-area surveyed, indicating that these are structural properties of the blockchain-security marriage rather than artifacts of any single application domain.
\item An explicit, evidence-based answer (Section~\ref{sec:timeliness}) to the question of whether this SoK is timely in 2026, together with a concrete research agenda (Section~\ref{sec:future}) for the most underexplored cell of the taxonomy: blockchain-backed EDR/XDR with automated, smart-contract-driven response.
\end{itemize}

The remainder of this paper is organized as follows. Section~\ref{sec:background} provides background on IDS/IPS/EDR taxonomy and the blockchain primitives most frequently borrowed by the security literature. Section~\ref{sec:methodology} describes the systematization methodology. Section~\ref{sec:taxonomy} presents the three-axis taxonomy. Sections~\ref{sec:ids-ips} and~\ref{sec:edr} synthesize the IDS/IPS and EDR/XDR literature respectively. Section~\ref{sec:crosscutting} discusses cross-cutting open issues. Section~\ref{sec:timeliness} addresses timeliness directly. Section~\ref{sec:future} lays out future research directions, and finally, Section~\ref{sec:conclusion} concludes this paper.

\section{Background}
\label{sec:background}

\subsection{Detection-and-Response System Classes}

We adopt the following working definitions, consistent with usage across the surveyed literature.

\textbf{Network Intrusion Detection System (NIDS):} Monitors network traffic (packet headers, flow statistics) for malicious activity, typically deployed at chokepoints such as gateways or routers.

\textbf{Host Intrusion Detection System (HIDS):} Monitors a single host's internal state (e.g., system calls, process behavior, log files) to detect anomalies. Embedded and industrial variants of HIDS have been proposed for resource-constrained devices~\cite{martinez2021host}, and feature-engineering approaches such as TF-IDF-vectorized system-call analysis have been used to detect anomalous processes~\cite{subba2021tfidfvectorizer}.

\textbf{Intrusion Prevention System (IPS):} Extends detection with active mitigation (e.g., dropping packets, blocking IPs, or quarantining hosts) ideally in real time. Despite the widespread use of the term IDS/IPS, the surveyed literature is dominated by detection-oriented systems, with comparatively few proposals providing genuine prevention functionality~\cite{miller2024collaborative}.

\textbf{Endpoint Detection and Response (EDR):} A more recent commercial and research category emphasizing continuous endpoint telemetry collection, behavioral analytics, automated investigation, and orchestrated response (isolation, rollback, remediation) at the individual endpoint, frequently cloud-coordinated across a fleet~\cite{kaur2024evolution}. EDR is distinguished from classical HIDS mainly by its emphasis on automated response and forensic tooling rather than detection alone~\cite{eigner2026comparison}.

\textbf{Extended Detection and Response (XDR):} Extends EDR by correlating telemetry from endpoints, networks, cloud workloads, identity services, and other security controls into a unified detection-and-response platform. Despite the growing adoption of XDR in operational cybersecurity, it receives little attention in the blockchain literature surveyed in this work, with most studies remaining focused on standalone IDS or EDR architectures rather than cross-domain detection and response.

\subsection{Blockchain Primitives Relevant to Detection-and-Response Design}\label{subsec:BlockchainPrimitives}
Rather than introducing fundamentally new blockchain mechanisms, the detection-and-response literature repeatedly builds upon a relatively small set of blockchain primitives. Existing proposals differ primarily in how these primitives are combined and integrated into security architectures, rather than in the primitives themselves. In other words, the detection-and-response literature borrows a relatively small set of blockchain primitives, repeatedly recombined:

%%%%%%%%%%%%%%%%%%%%%%%%%%%%%%%%%%%%%%%%
\textit{Immutable ledgers for evidentiary logging:} Detection metadata, alerts, and encrypted feature vectors can be committed to a ledger to guarantee tamper-evidence and traceability of intrusion alerts~\cite{chan2023complete,rahul2025blockchain}.

\textit{Consensus protocols for decentralized trust:} Rather than trusting a single coordinator, distributed detection nodes reach agreement via consensus ranging from classical Byzantine fault-tolerant protocols to novel domain-specific protocols such as proof-of-federated-quality consensus for decentralized threat hunting~\cite{hieu2025pofq} and multi-proof-of-work schemes tailored to constrained AIoT transaction validation~\cite{ar2025enhancing}.

\textit{Smart contracts for policy enforcement and automated response:} Self-executing code triggers actions when predefined conditions are met~\cite{liu2025blockchain}; in a detection-and-response context this enables automatic compliance enforcement and adaptive defense-mechanism adjustment~\cite{alevizos2025automated}, or constrains harmful behavior directly, as in threshold-based smart contracts that mitigate distributed denial-of-service (DDoS) conditions by limiting validator participation.

\textit{Federated learning integrity assurance:} Blockchain has been used specifically to make the model-update exchange step of federated-learning-based IDS verifiable and tamper-resistant, addressing the integrity gap left open by purely federated (non-blockchain) IDS architectures~\cite{ali2025federated}.

\textit{Incentive layers:} Token-based or reward-based mechanisms encourage honest participation in distributed detection tasks, as conceptualized in reward-driven IDS designs~\cite{selimi2023cybernfts}.

%These five primitives map directly onto the four functional roles introduced in Section~\ref{sec:intro} and form the second axis of the taxonomy developed in Section~\ref{sec:taxonomy}.

Four of these five primitives map one-to-one onto the four functional roles
introduced in Section I and used as Axis~2 of the taxonomy in Section~\ref{sec:taxonomy}: immutable ledgers realize the logging role~(L), consensus protocols realize the
trust role~(T), incentive layers realize the incentive role~(I), and
federated-learning integrity assurance realizes the FL-integrity role~(F). The
fifth primitive, smart contracts, is not assigned its own Axis-2 role; instead we
treat it as the cross-cutting implementation mechanism for Axis~3
(response-automation maturity), since a single contract can be used to
operationalize logging, consensus, incentive payout, or FL-update verification
depending on which role it is attached to. Table~\ref{tab:primitive-role-mapping}
makes this mapping explicit.

\begin{table}[t]
\centering
\caption{Mapping of Blockchain Primitives (Sec.~\ref{subsec:BlockchainPrimitives}) to Taxonomy Axes (Sec.~\ref{sec:taxonomy})}
\label{tab:primitive-role-mapping}
\begin{tabular}{ll}
\toprule
Primitive & Taxonomy mapping \\
\midrule
Immutable ledgers & Axis 2: Role L (evidentiary/audit) \\
Consensus protocols & Axis 2: Role T (trust/consensus) \\
Incentive layers & Axis 2: Role I (incentive) \\
FL integrity assurance & Axis 2: Role F (FL-integrity) \\
Smart contracts & Cross-cutting: implements Axis 3 \\
 & (R2/R3 response automation) atop L/T/I/F \\
\bottomrule
\end{tabular}
\end{table}

\begin{figure*}[t]
\centering
\begin{tikzpicture}[
    font=\sffamily\scriptsize,
    axisarrow/.style={-{Stealth[length=2.5mm]}, thick},
    box/.style={draw, rounded corners=2pt, minimum height=0.75cm, align=center},
    scale=0.85, every node/.append style={scale=0.85}
]

\node[draw, thick, rounded corners=4pt, fill=gray!5, minimum width=9.8cm, minimum height=4.8cm] (space) at (7.3, 2.1) {};

\node[box, fill=red!15, minimum width=3.2cm] (r0) at (0, 0.3) {R0: Detection only};
\node[box, fill=orange!20, minimum width=3.2cm] (r1) at (0, 1.5) {R1: Logged alerting};
\node[box, fill=yellow!30, minimum width=3.2cm] (r2) at (0, 2.7) {R2: Triggered mitigation};
\node[box, fill=green!25, minimum width=3.2cm] (r3) at (0, 3.9) {R3: Closed-loop orchestration};
\node[rotate=90, font=\bfseries\scriptsize] at (-2.2, 2.1) {Axis 3: Response Maturity};

\draw[axisarrow] (2.1, 5.2) -- (12.5, 5.2);
\node[box, fill=blue!10, minimum width=1.6cm] (nids) at (3.5, 5.2) {NIDS};
\node[box, fill=blue!10, minimum width=1.6cm] (hids) at (5.4, 5.2) {HIDS};
\node[box, fill=blue!10, minimum width=1.6cm] (ips)  at (7.3, 5.2) {IPS};
\node[box, fill=purple!15, minimum width=1.6cm] (edr) at (9.2, 5.2) {EDR};
\node[box, fill=purple!15, minimum width=1.6cm] (xdr) at (11.1, 5.2) {XDR};
\node[font=\bfseries\scriptsize] at (7.3, 5.9) {Axis 1: Detection-System Class};

\node[circle, draw, fill=yellow!60, minimum size=4.5pt, inner sep=0pt] (p1) at (3.2, 2.9) {};
\node[font=\tiny, anchor=west] at (3.3, 2.9) {IIoT-FL-IPS};

\node[circle, draw, fill=orange!60, minimum size=4.5pt, inner sep=0pt] (p2) at (3.2, 1.5) {};
\node[font=\tiny, anchor=west] at (3.3, 1.5) {Vehicular FL-IDS};

\node[circle, draw, fill=orange!60, minimum size=4.5pt, inner sep=0pt] (p3) at (4.5, 1.1) {};
\node[font=\tiny, anchor=west] at (4.6, 1.1) {UAV CGAN-IDS};

\node[circle, draw, fill=orange!60, minimum size=4.5pt, inner sep=0pt] (p4) at (5.6, 1.5) {};
\node[font=\tiny, anchor=west] at (5.7, 1.5) {PoFQ};

\node[circle, draw, fill=red!60, minimum size=4.5pt, inner sep=0pt] (p5) at (6.4, 0.4) {};
\node[font=\tiny, anchor=west] at (6.5, 0.4) {CyberNFTs};

\node[circle, draw, fill=yellow!60, minimum size=4.5pt, inner sep=0pt] (p6) at (6.6, 2.7) {};
\node[font=\tiny, anchor=west] at (6.7, 2.7) {AIoT multi-PoW};

\node[circle, draw, fill=green!60, minimum size=5.5pt, inner sep=0pt] (p7) at (9.1, 3.9) {};
\node[font=\tiny, anchor=west] at (9.2, 3.9) {Compliance-BC};

\node[circle, draw, fill=red!30, minimum size=4.5pt, inner sep=0pt] (p8) at (8.6, 0.4) {};
\node[font=\tiny, anchor=west] at (8.7, 0.4) {EDR review};

\node[draw=red, thick, dashed, rounded corners=3pt, fill=red!5, fill opacity=0.4, text opacity=1, minimum width=3.4cm, minimum height=2.4cm] (gap) at (10.4, 3.3) {};
\node[red, font=\bfseries\tiny, align=center] at (10.4, 3.3) {Largely empty:\\genuine\\per-endpoint\\R2/R3};

\end{tikzpicture}
\caption{Schematic view of the three-axis taxonomy (Section~\ref{sec:taxonomy}). Points are representative systems from Table~\ref{tab:taxonomy}, positioned by detection-system class (x-axis) and response-automation maturity (y-axis); point color/fill loosely indicates dominant blockchain functional role. The dashed region marks the largely unoccupied EDR/XDR $\times$ R2/R3 cell that motivates Sections~VI and~IX.}
\label{fig:taxonomy-schematic}
\end{figure*}
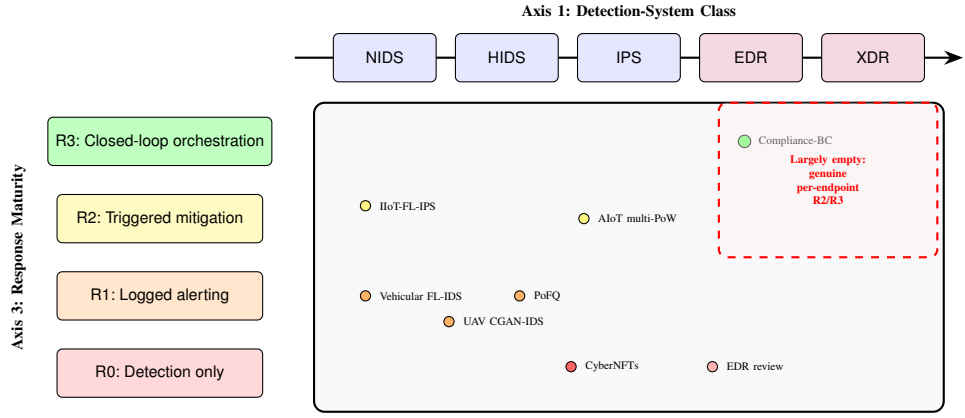

\section{Systematization Methodology}\label{sec:methodology}
Consistent with SoK methodology in the security literature, this paper does not propose a new detection algorithm; it systematizes existing knowledge to expose structure, redundancy, and gaps that are invisible when papers are read individually.

\subsection{Scope and Search Strategy}
We targeted high-impact venues, prioritizing IEEE Transactions and Journals (e.g., Internet of Things Journal, Transactions on Vehicular Technology, Transactions on Knowledge and Data Engineering, Transactions on Parallel and Distributed Systems, Transactions on Network and Service Management, Communications Surveys \& Tutorials, Access, etc.), ACM venues (e.g., Computing Surveys), Elsevier venues (e.g., Computers \& Security, Journal of King Saud University Computer and Information Sciences, etc.), Springer Nature venues (e.g., Scientific Reports, Journal of Cloud Computing, Journal of Big Data, Cluster Computing, etc.), and MDPI venues (e.g., Future Internet, Smart Cities, Applied Sciences, etc.), supplemented by arXiv preprints for the most recent (2025--2026) work not yet indexed in journal form. We included arXiv preprints because excluding them would omit the most recent LLM-blockchain and post-quantum-blockchain work central to Sections~\ref{sec:ai-engine} and~\ref{Post-QuantumCryptographic}. 
Search terms combined \{blockchain, distributed ledger, smart contract\} with \{intrusion detection, intrusion prevention, EDR, endpoint detection and response, federated learning, IDS\}.

\subsection{Inclusion Criteria}
A paper was included if it (i) proposed, evaluated, or systematically reviewed a security architecture in which blockchain or a distributed ledger plays a structural role (not merely cited as related background), and (ii) the architecture's primary function was intrusion/anomaly detection, intrusion prevention, or endpoint/extended detection and response. Pure blockchain-security papers (e.g., smart contract vulnerability detection without an IDS/EDR framing) were included only where directly relevant to the response-automation axis.

In addition to the blockchain-structural corpus defined above, we include a small set of $N_b$ \emph{non-blockchain reference systems} (marked~$\dagger$ in Table~I) solely to anchor the detection-class and response-maturity baselines discussed in Section~VI-A. These reference systems are excluded from all Axis-2
 blockchain functional role) statistics and from any claim of corpus completeness; they answer "what does R2/R3 look like without blockchain," not "how mature is blockchain-EDR."

\subsection{Why a Taxonomy Rather Than a Chronological Catalogue}
Existing systematic reviews in this space are predominantly organized chronologically or by application domain (IoT, IIoT, vehicular, healthcare)~\cite{hassan2023survey,adele2024comprehensive}. This is useful for breadth but obscures the functional-role conflation described in Section~\ref{sec:intro}: two IoT-domain papers may use blockchain for entirely different purposes (one for tamper-evident logging, another for FL integrity) yet appear adjacent in a domain-organized table. We instead organize around \emph{function}, which is what determines which open problems (Section~\ref{sec:crosscutting}) actually apply to a given system.

\section{A Three-Axis Taxonomy}\label{sec:taxonomy}
We propose classifying each surveyed system along three independent axes, summarized in Table~\ref{tab:taxonomy}.

\textbf{Axis 1 -- Detection-system class:} NIDS, HIDS, IPS (detection + active mitigation), EDR, or XDR, as defined in Section~\ref{sec:background}.

\textbf{Axis 2 -- Blockchain functional role.} Evidentiary/audit logging (L), decentralized trust/consensus (T), incentive mechanism (I), or FL-integrity assurance (F). A system may exploit more than one role simultaneously.

\textbf{Axis 3 -- Response-automation maturity:} We define a four-level ordinal scale:
\begin{itemize}
\item \textit{R0 -- Detection only:} The system flags anomalies; no automated action follows.
\item \textit{R1 -- Logged alerting:} Detections are committed to the ledger for audit purposes, but mitigation is manual.
\item \textit{R2 -- Triggered mitigation:} A smart contract or off-chain agent automatically executes a bounded action (e.g., isolate node, drop traffic, revoke credential) upon detection.
\item \textit{R3 -- Closed-loop orchestration:} Detection, response, and verification of remediation are integrated into a single auditable pipeline with feedback to the detection model (the EDR/XDR ideal).
\end{itemize}

\begin{table*}[t]
\centering
\caption{Three-Axis Taxonomy Applied to Representative Surveyed Systems}
\label{tab:taxonomy}
\begin{tabular}{p{0.30\textwidth}p{0.12\textwidth}p{0.10\textwidth}p{0.08\textwidth}p{0.30\textwidth}}
\toprule
\textbf{Representative System / Source} & \textbf{Detection Class} & \textbf{BC Role} & \textbf{Response} & \textbf{Notes} \\
\midrule
Lightweight federated intrusion prevention, IIoT~\cite{nishad2026lightweight} & IPS (IIoT) & T, F & R2 & Resource-constrained devices; scalability bottleneck reported \\
Blockchain-FL collaborative IDS, vehicular edge~\cite{liu2021blockchain} & NIDS (vehicular) & T, F & R1 & IEEE TVT; FL aggregation secured on-chain \\
CGAN-based collaborative IDS, UAV networks~\cite{he2022cgan} & NIDS (UAV) & T, F & R1 & Distributed FL across UAV swarm \\
PoFQ consensus for decentralized threat hunting~\cite{hieu2025pofq} & NIDS & T & R1--R2 & Novel consensus protocol, trustless setting \\
Securing FL with blockchain + XAI, IoT~\cite{abou2023securing} & NIDS (IoT) & F & R1 & Explainability layer added to FL-IDS \\
CyberNFTs reward-driven IDS~\cite{selimi2023cybernfts} & NIDS & I, T & R0--R1 & Conceptual; incentive layer is the core contribution \\
Smart-contract DDoS-mitigating PKI & NIDS/IPS & T & R2 & Threshold smart contract blocks DDoS at validator layer \\
Anomaly detection + multi-PoW consensus, AIoT~\cite{ar2025enhancing} & NIDS (AIoT) & T, L & R2 & Smart contracts automate device registration/response \\
Automated compliance \& threat response (AI+BC+SC)~\cite{alevizos2025automated} & EDR-adjacent & T, L & R3 & Closest reviewed system to closed-loop EDR; not endpoint-specific \\
EDR evolution review~\cite{kaur2024evolution} & EDR & (peripheral) & R0--R1 typical & Blockchain treated as data-security adjunct, not structural \\
CrowdStrike/SentinelOne comparison~\cite{eigner2026comparison} & EDR (commercial) & none & R2--R3 (proprietary) & No blockchain integration; included as EDR-maturity baseline \\
Host IDS for embedded industrial devices~\cite{martinez2021host} & HIDS & none & R0 & Pre-blockchain baseline; included for contrast \\
LLM detector + post-quantum blockchain, IoT (LQB-IDS)~\cite{huang2025adaptive} & NIDS (IoT) & T, L & R1 & Closest reviewed system combining a modern AI engine with blockchain \emph{and} post-quantum security in one design \\
\bottomrule
\end{tabular}
\end{table*}

Table~\ref{tab:taxonomy} makes the central finding of this SoK visually explicit: the T (trust/consensus) and F (FL-integrity) roles dominate the NIDS column; response automation rarely exceeds R1--R2; and the EDR column is populated almost entirely by systems with no blockchain integration at all, with the single closest analogue~\cite{alevizos2025automated} being compliance-and-policy-oriented rather than endpoint-telemetry-oriented. No surveyed system reaches R3 with a genuine per-endpoint blockchain-anchored response loop, which we revisit in Section~\ref{sec:future}.

\section{Blockchain-Based Network and Host IDS/IPS}\label{sec:ids-ips}
This section synthesizes the dominant body of literature (i.e., blockchain applied to network- and host-level detection) organized by the functional role taxonomy of Section~\ref{sec:taxonomy} rather than by application domain, to surface cross-domain patterns that domain-organized surveys obscure.

\subsection{Trust and Consensus for Collaborative Detection}
The earliest and most persistent use of blockchain in this literature is to replace a trusted central aggregator with a consensus mechanism among mutually distrusting detection nodes. Foundational architectural work framed this explicitly as a trust problem: Kolokotronis et al.\ examined blockchain architectures specifically for trust-based collaborative intrusion detection, motivated by the observation that traditional collaborative IDS require a degree of inter-organizational trust that is often unrealistic~\cite{kolokotronis2019blockchain}. This framing recurs throughout subsequent work. A large catalogue of blockchain-based collaborative IDS proposals (spanning energy systems, software-defined networking, healthcare, and general IoT) has since been compiled and shows intrusion detection as by far the most common tagged function, with intrusion prevention and response trailing well behind~\cite{miller2024collaborative}.

Consensus-protocol innovation specific to the detection use case is an active sub-thread. Beyond adapting classical Byzantine fault-tolerant or proof-of-work consensus, several proposals design detection-specific consensus: a blockchain consensus protocol explicitly designed for decentralized federated-learning-based threat hunting in a trustless computing landscape has been proposed under the name PoFQ~\cite{hieu2025pofq}, and multi-proof-of-work schemes have been used specifically to minimize resource usage while validating AIoT transactions during anomaly analysis~\cite{ar2025enhancing}. The recurring design pressure in all of these is the same: classical consensus (e.g., PBFT-style protocols) imposes coordination overhead that trades throughput for the integrity guarantees decentralization provides~\cite{ayadi2026blockchain}, a tension we return to in Section~\ref{sec:crosscutting}.

\subsection{Federated-Learning Integrity Assurance}
A dominant pattern, particularly in IEEE-indexed venues, pairs Federated Learning (FL) for distributed model training with blockchain for securing the model-update exchange step. This combination directly addresses two separate weaknesses simultaneously: FL's privacy benefit (raw data never leaves the local node) and blockchain's integrity guarantee (model updates cannot be silently tampered with in transit or at aggregation).

Representative IEEE-venue instances include a blockchain-and-FL collaborative intrusion detection architecture for vehicular edge computing published in IEEE Transactions on Vehicular Technology~\cite{liu2021blockchain}, and a CGAN-based collaborative intrusion detection scheme for UAV networks that combines blockchain-empowered distributed FL, published in IEEE Internet of Things Journal~\cite{he2022cgan}. A 2025 IEEE Transactions on Knowledge and Data Engineering contribution extends blockchain-based FL intrusion detection specifically for IoT settings~\cite{sun2024blockchain}. Closer to the privacy-preservation angle, FL combined with blockchain and explainable AI has been used to secure IoT intrusion detection while additionally surfacing model decisions to human analysts~\cite{abou2023securing}, and a recent taxonomy paper frames privacy and trust as the two central axes specifically for blockchain-federated IDS, cataloguing challenges along both dimensions~\cite{cao2025privacy}.

Industrial IoT (IIoT) is a particularly active sub-domain for this combination. A 2025 framework explicitly designed for IIoT integrates a permissioned blockchain with FL to ensure tamper-resistant recording of model updates and anomaly scores while keeping raw industrial telemetry local to the edge device~\cite{ali2025federated}, directly addressing the model-integrity gap left open by purely federated (non-blockchain) IIoT IDS architectures. A 2026 lightweight blockchain-enabled federated intrusion prevention framework targets resource-constrained industrial IoT devices specifically, reporting that while accuracy improvements were substantial, scalability and resource-utilization overhead remained significant obstacles on constrained hardware~\cite{nishad2026lightweight}. Domain-specific instances extend this pattern to the Internet of Medical Things, where blockchain-based FL has been used for intrusion detection while preserving patient-data privacy under regulatory constraints~\cite{zaabar2022intrusion}, and to smart-home IoT, where knowledge-distillation has been combined with blockchain-enabled FL to compress the on-device detection model without sacrificing the integrity guarantees blockchain provides~\cite{shalan2025enhancing}.

Broader survey-level treatments of blockchain-empowered FL (not always IDS-specific but frequently citing IDS as a primary use case) include an ACM Computing Surveys treatment of blockchain-enabled FL generally~\cite{qu2022blockchain}, a Future Internet survey specifically on blockchain-based FL~\cite{wu2023survey}, and IEEE Internet of Things Journal work on the opportunities and challenges of combining FL with blockchain at the network edge~\cite{nguyen2021federated}. Resource-allocation and performance-analysis work for blockchain-assisted decentralized FL, such as the BLADE-FL framework published in IEEE Transactions on Parallel and Distributed Systems, is frequently cited as foundational infrastructure underpinning the IDS-specific instantiations above~\cite{li2021blockchain}, as is asynchronous blockchain-empowered FL work for secure data sharing in the Internet of Vehicles (IoV)~\cite{lu2020blockchain}.

\subsection{Evidentiary and Audit Logging}

A second, more narrowly scoped role treats the blockchain purely as a tamper-evident store for detection metadata, independent of any FL component. Architectures in this category integrate detection metadata and encrypted feature vectors directly into blocks specifically to guarantee block immutability and end-to-end traceability of intrusion alerts in wireless sensor network deployments~\cite{elumalai2026blockchain}. This role is conceptually simpler than the FL-integrity role (it requires no distributed training infrastructure) and is correspondingly easier to retrofit onto an existing centralized detection model, which may explain its continued appearance in industrial and embedded contexts where full FL infrastructure is impractical.

\subsection{Incentive Mechanisms}

The least mature of the four functional roles, but conceptually distinct, is the use of token- or reward-based incentive layers to align participant behavior in distributed detection settings. The CyberNFTs proposal explicitly frames intrusion detection as a reward-driven, decentralized task, conceptualizing a system in which correctly identified intrusions are rewarded via non-fungible-token-based mechanisms layered atop a machine-learning detection core~\cite{selimi2023cybernfts}. This remains largely conceptual in the surveyed literature; we did not find a high-impact-venue instance that combines incentive mechanisms with a deployed, evaluated detection pipeline at the scale of the FL-integrity work surveyed above, marking this as both the least mature and most open functional role.

\subsection{Active Prevention and DDoS-Specific Mitigation}

A meaningful sub-literature focuses specifically on denial-of-service mitigation, where the temporal pressure of an active attack makes the detection/response distinction sharpest. Optimized deep-learning approaches embedded within blockchain technology have been proposed specifically for DDoS-attack prevention~\cite{ilyas2023prevention}, and blockchain-based approaches to credible link-flooding-attack detection and mitigation have been published in IEEE Transactions on Network and Service Management~\cite{jiang2024credible}. Smart-contract-level prevention is exemplified by threshold-based contracts within blockchain-assisted PKI systems, where a smart contract validates the number of nodes permitted to act as a certificate authority for any single transaction, thereby bounding the attack surface for DDoS at the validator layer itself rather than detecting an attack after the fact. A recent hierarchical analysis of DoS attacks and defense technologies specifically within blockchain systems treats the blockchain itself as the asset under attack and surveys defenses accordingly, offering a complementary perspective to the IDS-for-blockchain framing that dominates elsewhere in this corpus~\cite{zhang2025attacks}.

\subsection{AI- and LLM-Based Detection Engines Within Blockchain Architectures}
\label{sec:ai-engine}

The preceding subsections already show that nearly every blockchain-IDS/IPS system surveyed relies on an AI or ML detection engine underneath the blockchain trust layer---XGBoost~\cite{song2025hybrid}, deep neural networks~\cite{ar2025enhancing}, CGANs~\cite{he2022cgan}, and federated deep learning more generally~\cite{ali2025federated}. We treat this here as a separate, fourth implicit axis (\emph{detection-engine family}) because it cuts across the functional-role taxonomy of Section~\ref{sec:taxonomy} and is undergoing its own rapid shift toward large language models (LLMs) that warrants explicit discussion.

A growing 2025--2026 sub-literature applies LLMs directly to network intrusion detection, motivated by LLMs' ability to operate with little or no labeled training data and to produce human-readable rationale for a flagged event. Foundational and survey-level treatments include a comprehensive survey of LLM applications, vulnerabilities, and defense techniques in cybersecurity~\cite{jaffal2025large}, and a dedicated treatment of LLMs for network intrusion detection systems covering foundations, implementations, and future directions~\cite{yang2025large}. Architecturally novel proposals include a tri-LLM cooperative federated framework that performs zero-shot intrusion detection through semantic disagreement and trust-aware aggregation across multiple LLM agents~\cite{jamshidi2026tri}, illustrating that the same trust/consensus problem central to blockchain-IDS (Section~\ref{sec:ids-ips}.A) is now also being addressed, independently, at the level of multi-LLM agent agreement.

Most directly relevant to this SoK is recent work that combines an LLM detection engine with a blockchain trust layer explicitly, rather than treating the two as separate research threads. A 2026 framework, LQB-IDS, proposes an adaptive IoT intrusion detection system that pairs an LLM-based detector with a post-quantum-secure blockchain, explicitly addressing both the labeled-data dependency of conventional dynamic IoT IDS and the data-security concerns largely neglected in prior LLM-based IoT IDS work, while reporting strong detection performance without requiring labeled training data~\cite{huang2025adaptive}. This single system is, to our knowledge, the closest existing instance of a fully integrated AI-engine-plus-blockchain-plus-post-quantum architecture in the surveyed corpus, simultaneously touching three of the cross-cutting open issues discussed in Section~\ref{sec:crosscutting} (detection-engine modernization, blockchain trust, and post-quantum exposure) within one design.

A related but distinct thread uses blockchain not to secure an IDS, but to secure the LLM itself, which is conceptually the inverse of this SoK's framing and worth noting for completeness. BC4LLM proposes using blockchain to make LLM behavior more trustworthy and auditable~\cite{luo2023bc4llm}, and a broader holistic survey catalogs blockchain-for-LLM-security work generally, including LLM-assisted smart-contract vulnerability detection as a recurring use case~\cite{geren2025blockchain}. We distinguish this ``blockchain secures the LLM'' direction from the ``LLM-plus-blockchain secures the network/endpoint'' direction that is the primary subject of this paper, since conflating them would repeat the functional-role-conflation problem identified in Section~\ref{sec:intro} at a different layer of abstraction.

\textbf{Implication for the taxonomy:} We do not currently see LLM-based detection engines applied to EDR/XDR-class systems in the surveyed literature, despite EDR being, if anything, an even more natural fit: LLMs' strength in synthesizing heterogeneous, semi-structured telemetry (process trees, command-line arguments, log text) maps closely onto exactly the kind of endpoint data EDR systems collect. The absence of LLM-based EDR-blockchain work is consistent with, and arguably reinforces, the broader EDR-blockchain gap identified in Section~\ref{sec:edr}, and we add it explicitly to the future-work agenda in Section~\ref{sec:future}.

It is necessary to be explicit about prior systematization efforts to justify the present contribution. A systematic literature review focused specifically on blockchain-based IDS/IPS for IoT networks reviewed multiple relevant proposals and their underlying mechanisms~\cite{hassan2023survey}. A broader and more recent systematic review examined 113 articles published across IPS, IDS, and IDRS research over the preceding seven years, providing an up-to-date survey of the field as of its publication~\cite{adele2024comprehensive}. A literature review on adapting security and decentralized knowledge enhancement in FL via blockchain technology surveys blockchain-FL combinations broadly, with IDS as one of several application areas~\cite{orabi2025adapting}. Each of these is organized chronologically or by application domain. None separates the four functional roles identified in Section~\ref{sec:background}, and none extends its scope to EDR/XDR, which is the gap addressed in Section~\ref{sec:edr}.

\section{Blockchain and Endpoint Detection and Response: A Gap Analysis}
\label{sec:edr}

\subsection{The State of EDR Research Independent of Blockchain}

EDR as a research and commercial category has matured substantially on its own terms. A comprehensive review of EDR's evolution in cybersecurity highlights AI and machine learning as the dominant transformative forces in threat detection and response, with blockchain mentioned specifically for its contribution to securing digital transactions and data through decentralization, transparency, and immutability---positioned alongside AI/ML rather than as a load-bearing architectural component~\cite{kaur2024evolution}. The same review explicitly discusses decentralized architectures as part of EDR's anticipated future direction, but does not develop a blockchain-specific detection-and-response architecture~\cite{kaur2024evolution}. Separately, a detailed empirical comparison of two leading commercial EDR platforms, CrowdStrike and SentinelOne, found that platform suitability depends heavily on deployment context---one platform proving more suitable for decentralized or frequently offline environments and organizations preferring stronger autonomous protection and automated remediation~\cite{eigner2026comparison}. Notably, ``decentralized'' here refers to organizational/network topology, not to blockchain-based architectural decentralization---a terminological overlap that is itself worth flagging for researchers entering this space, since the two senses of ``decentralized'' are easy to conflate when reading across the EDR and blockchain-security literature.

\subsection{Where Blockchain Appears Adjacent to EDR, but Not Within It}

A small number of papers sit at the boundary of EDR and blockchain without fully occupying the intersection. An automated cybersecurity compliance and threat-response framework integrates AI, blockchain, and smart contracts to automate the enforcement of security policies and accelerate threat response, using blockchain specifically for transparent, immutable logging of compliance actions and smart contracts to ensure uniform application of security measures. This is architecturally the closest analogue to a blockchain-EDR system found in the surveyed corpus (it exhibits R3-level closed-loop characteristics in our taxonomy) but its unit of action is an organizational security policy rather than an individual endpoint's telemetry stream, and it is not benchmarked against endpoint-level detection workloads.

A second boundary case is integrity-focused rather than detection-focused: a 2026 framework combining AI-assisted vulnerability discovery with a blockchain-backed integrity layer reports evaluation on a dataset of 500 endpoints, achieving substantial detection accuracy with only moderate computational overhead, explicitly demonstrating that decentralized integrity assurance can coexist with high-speed security workflows~\cite{parmar2026zer0n}. This is the single closest match to a per-endpoint blockchain-secured detection pipeline identified in our search, yet its framing is vulnerability discovery rather than runtime behavioral detection-and-response, and it does not implement the orchestrated remediation (isolation, rollback) that defines EDR as distinct from vulnerability scanning.

\subsection{Why This Gap Likely Exists}

We offer three non-exclusive explanations for the EDR-blockchain gap, intended as hypotheses for future empirical or qualitative investigation rather than established fact.

First, \emph{latency mismatch}. EDR's value proposition rests on rapid (often sub-second) automated containment of an active compromise. Consensus finality in most blockchain architectures used in the surveyed IDS literature is measured in seconds at best, and the literature consistently reports scalability and resource-utilization difficulties even in the comparatively latency-tolerant IoT-IDS setting~\cite{nishad2026lightweight}. An EDR system that must wait on multi-node consensus before isolating a compromised endpoint may simply be too slow to be useful, which would explain why none of the surveyed systems push response automation to R3 at the individual-endpoint level.

Second, \emph{deployment-model mismatch}. Commercial EDR is overwhelmingly vendor-centralized. A single vendor's cloud back-end correlates telemetry across a customer's fleet~\cite{eigner2026comparison}. Blockchain's value proposition (removing a trusted central party) is least compelling precisely where a trusted central vendor relationship already exists by design, which may reduce the perceived incentive for blockchain integration relative to the genuinely multi-stakeholder, no-central-authority settings (cross-organization IoT, vehicular networks, federated consortia) where blockchain-IDS research has concentrated.

Third, \emph{research-community separation}. EDR research output in the surveyed literature originates largely from industry-adjacent and applied-security venues, while blockchain-IDS research is concentrated in IEEE/ACM/Elsevier academic venues with a strong FL and IoT-systems orientation. The two communities appear to read and cite each other minimally, based on the near-total absence of cross-citation observed in this search.

\subsection{Implications}

Whichever explanation (or combination) is correct, the practical implication is the same: a credible blockchain-EDR architecture cannot simply port the FL-integrity or trust-consensus patterns dominant in Section~\ref{sec:ids-ips} onto an endpoint-telemetry setting without first resolving the latency mismatch. This motivates the layered/hybrid architectural direction proposed in Section~\ref{sec:future}.
%%%%%%%%%%%%%%%%%%%%%%%%%%%%%%%%%%%%%%%%%%5
\section{Cross-Cutting Open Issues}
\label{sec:crosscutting}

This section synthesizes problems that recur across nearly every functional role and detection class surveyed, indicating that they are structural to the blockchain-security pairing rather than artifacts of any single sub-domain.

\subsection{Scalability and Consensus Latency}

This is the single most frequently reported limitation in the corpus. Direct empirical testing of blockchain-based IDS/IPS in IoT networks has surfaced significant scalability and resource-utilization issues, particularly on limited-capability systems~\cite{nishad2026lightweight}. The same source reports that while blockchain-machine-learning integration (e.g., XGBoost-based detection) can push detection accuracy scores extremely high, genuine real-time processing and the associated computational overhead remain a problem in practice~\cite{nishad2026lightweight}. Federated learning compounds rather than resolves this: collaborative intrusion detection studies in resource-constrained IoT contexts have observed that microcontrollers and other low-capability devices struggle specifically with the federated learning component, independent of any blockchain overhead layered on top~\cite{ieropoulos2026empowering}. At the consensus layer specifically, permissioned blockchain deployments using strong Byzantine-fault-tolerant-style consensus protocols generally exhibit reduced throughput compared to optimized centralized databases, since consensus validation imposes extra coordination overhead as a direct trade-off against the decentralized integrity guarantees being purchased~\cite{ayadi2026blockchain}.

This overhead is not a peripheral implementation detail; it directly determines which axis-3 response-automation level (Section~\ref{sec:taxonomy}) a system can realistically target. Systems requiring R2 (triggered mitigation) or R3 (closed-loop orchestration) respond automatically must complete consensus within a window short enough to be operationally useful, which the current generation of general-purpose consensus protocols struggles to guarantee under the device constraints typical of IoT and IIoT deployments.

\subsection{Post-Quantum Cryptographic Exposure}\label{Post-QuantumCryptographic}

A separate but increasingly cited cross-cutting concern is the long-term cryptographic exposure of blockchain-based security architectures to quantum computing advances. The cryptographic foundations of blockchain (i.e., hashing and digital signatures in particular) are directly threatened by quantum algorithms, specifically Shor's and Grover's algorithms, which compromise the hardness assumptions underlying current public-key schemes~\cite{gharavi2024post}. Two broad mitigation approaches have been surveyed: post-quantum blockchains, which substitute classical-but-quantum-resistant cryptographic algorithms into existing blockchain designs, and quantum blockchains, which use quantum-mechanical properties directly~\cite{yang2023survey}. A systematic survey of post-quantum distributed ledger technology frames this transition as both urgent and currently under-addressed in deployed systems, aiming to give researchers and developers a practical, comprehensive perspective on post-quantum blockchain security specifically to stimulate further work at this intersection~\cite{baseri2025blockchain}. IoT-specific treatments argue that securing IoT in the post-quantum era requires attention to implementation feasibility on constrained hardware, not merely theoretical cryptographic soundness~\cite{khan2025securing}.

None of the blockchain-IDS/IPS or blockchain-EDR-adjacent systems surveyed in Sections~\ref{sec:ids-ips}--\ref{sec:edr} discuss post-quantum readiness explicitly, despite many being designed for long-lived industrial or vehicular deployments where cryptographic agility over a multi-year or multi-decade hardware lifecycle is a realistic operational concern. This represents a clear and currently unaddressed cross-cutting gap.

\subsection{Smart-Contract Attack Surface}

Using smart contracts for automated response (Section~\ref{sec:ids-ips}, R2/R3 systems) introduces the smart contract itself as a new attack surface. A lifecycle-perspective survey of smart-contract security catalogues a substantial range of threats and prevention-stage mitigation strategies, including balance-verification mechanisms to prevent reentrancy and bytecode-obfuscation tools to resist reverse engineering~\cite{liu2025blockchain}. A response-automation architecture that relies on a vulnerable smart contract to trigger endpoint isolation or traffic blocking inherits this attack surface directly; an adversary capable of exploiting the response contract could in principle weaponize the very mechanism intended to contain them. This risk is essentially unaddressed in the IDS/IPS-specific literature surveyed in Section~\ref{sec:ids-ips}, which tends to treat the smart-contract layer as a trusted given rather than as a component requiring its own security analysis.

\subsection{Adversarial Robustness of the AI Detection Engine}

A further cross-cutting issue, distinct from but related to the smart-contract attack surface above, concerns the AI/ML detection engine itself, which nearly every surveyed system treats as a trusted oracle whose output the blockchain layer faithfully logs or acts upon. Blockchain's integrity guarantee covers the immutability of a recorded detection decision; it says nothing about whether that decision was correct, and a detection model that has been poisoned, evaded, or otherwise manipulated will have its incorrect verdict immutably and verifiably recorded with the same confidence as a correct one. Federated-learning-based detection pipelines are particularly exposed here, since a malicious participant can submit poisoned model updates; blockchain-based monitoring techniques that decouple a monitoring phase from a detection phase specifically to catch targeted and untargeted poisoning attacks in decentralized FL have been proposed as a partial mitigation~\cite{al2022blockchain}, but this remains a minority practice rather than a default design assumption across the corpus.

The shift toward LLM-based detection engines (Section~\ref{sec:ai-engine}) introduces an additional, less-studied attack surface: prompt-injection-style manipulation of the LLM's reasoning process, and the well-documented tendency of LLMs toward hallucination, both of which could cause an LLM-based detector to misclassify traffic or endpoint behavior in ways that differ qualitatively from the failure modes of classical ML detectors. None of the LLM-blockchain-IDS work surveyed in Section~\ref{sec:ai-engine}, including LQB-IDS~\cite{huang2025adaptive}, evaluates robustness against adversarially crafted inputs designed to exploit the LLM component specifically, leaving this as a concrete, currently open gap at the intersection of LLM safety research and blockchain-IDS design.

\subsection{Evaluation Methodology Weaknesses}

A recurring pattern across the corpus is evaluation against synthetic or simulated workloads rather than production-representative traffic, and inconsistent baselines across papers claiming similar accuracy improvements. Multiple surveyed systems report very high (often $>$99\%) detection-accuracy figures while separately acknowledging substantial real-time processing overhead~\cite{nishad2026lightweight}, raising the question of whether reported accuracy figures are obtained under conditions representative of the latency constraints the same papers identify as a limitation. Systematic reviews of the broader IDS/IPS/IDRS literature note this tension only implicitly, by cataloguing methodology diversity rather than critiquing it directly~\cite{adele2024comprehensive}. We flag evaluation-methodology standardization (common datasets, common latency-under-load reporting, and consistent accuracy/overhead trade-off curves rather than point estimates) as a cross-cutting open issue that would materially improve comparability across this literature.

\subsection{Interoperability Across Heterogeneous Blockchain Deployments}

Finally, most surveyed systems assume a single, homogeneous blockchain deployment shared by all participating nodes. Real-world collaborative detection, particularly across organizational boundaries (multiple hospitals, multiple fleet operators, multiple critical-infrastructure owners), is more likely to involve heterogeneous, independently governed ledgers that must interoperate. This cross-chain interoperability problem is well studied in the general blockchain literature but is essentially absent from the IDS/IPS/EDR-specific corpus surveyed here, representing a further open issue at the intersection of the two fields.

\section{Is This SoK Timely in 2026?}
\label{sec:timeliness}

We return directly to the question posed at the outset, since it deserves an explicit, evidence-based answer rather than an assumed one.

\textbf{The case against timeliness} is straightforward: blockchain-IDS-for-IoT is a saturated subfield. Several systematic reviews already exist (e.g.,~\cite {adele2024comprehensive}), a dedicated IoT-specific SLR already exists~\cite{hassan2023survey}, and the publication rate in this specific niche (blockchain + FL + IoT-IDS) shows no sign of slowing, with multiple 2025--2026 contributions identified in Section~\ref{sec:ids-ips} alone. A SoK that re-surveys this ground without a differentiated angle would add little.

\textbf{The case for timeliness, conditional on scope}, rests on four points established in this paper. First, the EDR/XDR-blockchain intersection (Section~\ref{sec:edr}) is demonstrably underexplored relative to its IoT-IDS counterpart. We found essentially no high-impact-venue paper that builds a genuine per-endpoint blockchain-anchored detection-and-response pipeline, as distinct from policy-compliance automation~\cite{alevizos2025automated} or vulnerability-discovery integrity assurance~\cite{parmar2026zer0n}. A systematization that names this gap precisely and explains plausible causes (Section~\ref{sec:edr}-D) has not previously been published. Second, the cross-cutting problems identified in Section~\ref{sec:crosscutting} (particularly post-quantum exposure) have only recently matured into citable, venue-published concerns (IEEE Communications Surveys \& Tutorials 2024~\cite{yang2023survey}; multiple 2025--2026 post-quantum-blockchain contributions) and have not yet been connected to the IDS/IPS/EDR literature at all; doing so now is timely precisely because the connection has not yet been made, not because the underlying post-quantum literature is itself new. Third, the shift toward LLM-based detection engines (Section~\ref{sec:ai-engine}) is itself extremely recent. For instance, the LQB-IDS system combining an LLM detector with post-quantum blockchain was published in early 2025~\cite{huang2025adaptive}. This means a systematization connecting this emerging detection-engine family to the blockchain-trust and response-automation axes developed here could not have been written meaningfully even one to two years earlier. Fourth, the functional-role taxonomy proposed in Section~\ref{sec:taxonomy}, separating logging, trust/consensus, incentive, and FL-integrity roles, is a contribution independent of any single publication date. It is a structural reframing of existing knowledge, which is precisely what distinguishes an SoK from a literature review and is unaffected by whether the underlying corpus continues to grow.

Our conclusion is that a 2026 SoK on this topic makes sense, but only under the EDR/XDR-inclusive, function-separated framing developed here---not as a re-statement of the existing IoT-IDS survey literature.

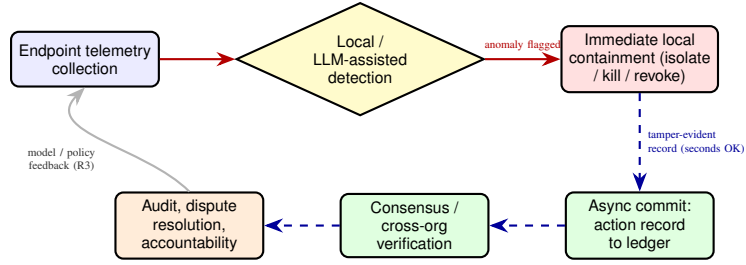
\begin{figure*}[t]
\centering
\begin{tikzpicture}[
    font=\sffamily\scriptsize,
    node distance=0.85cm and 1.0cm,
    box/.style={draw, thick, rounded corners=3pt, minimum height=0.85cm, minimum width=2.3cm, align=center, fill=blue!8},
    dec/.style={draw, thick, diamond, aspect=2.2, align=center, fill=yellow!25, minimum width=2.1cm, minimum height=0.85cm},
    ledgerbox/.style={draw, thick, rounded corners=3pt, minimum height=0.85cm, minimum width=2.3cm, align=center, fill=green!12},
    arr/.style={-{Stealth[length=2.5mm]}, thick},
    fastarr/.style={-{Stealth[length=2.5mm]}, thick, red!70!black},
    slowarr/.style={-{Stealth[length=2.5mm]}, thick, dashed, blue!60!black},
    scale=0.85, every node/.append style={scale=0.85}
]

\node[box] (telemetry) {Endpoint telemetry\\collection};
\node[dec, right=of telemetry] (detect) {Local /\\LLM-assisted\\detection};
\node[box, right=of detect, fill=red!12] (contain) {Immediate local\\containment (isolate\\/ kill / revoke)};

\draw[fastarr] (telemetry) -- (detect);
\draw[fastarr] (detect) -- node[above, font=\tiny, red!70!black] {anomaly flagged} (contain);

\node[ledgerbox, below=1.3cm of contain] (log) {Async commit:\\action record\\to ledger};
\node[ledgerbox, left=of log] (consensus) {Consensus /\\cross-org\\verification};
\node[ledgerbox, left=of consensus, fill=orange!15] (audit) {Audit, dispute\\resolution,\\accountability};

\draw[slowarr] (contain) -- node[right, font=\tiny, blue!60!black, align=left] {tamper-evident\\record (seconds OK)} (log);
\draw[slowarr] (log) -- (consensus);
\draw[slowarr] (consensus) -- (audit);

\draw[arr, gray!60] (audit.north) to[out=135, in=225, looseness=1.1] node[below left, font=\tiny, gray!40!black, pos=0.5, align=center] {model / policy\\feedback (R3)} (telemetry.south);

\end{tikzpicture}
\caption{Proposed hybrid on-chain/off-chain response architecture (Section~IX.A). The fast path (solid red) executes containment locally within an EDR-realistic latency budget, without waiting on consensus. The slow path (dashed blue) asynchronously commits a tamper-evident audit record to the ledger, where consensus latency is acceptable. The optional feedback loop (gray) represents the R3 closed-loop ideal, which no surveyed system currently implements at the individual-endpoint level.}
\label{fig:hybrid-architecture}
\end{figure*}

\section{Future Research Directions}
\label{sec:future}

Building directly on the gaps identified in Sections~\ref{sec:edr} and~\ref{sec:crosscutting}, we propose the following concrete research directions, ordered roughly by tractability.

\subsection{Hybrid On-Chain/Off-Chain Response Architectures}

To resolve the latency mismatch identified in Section~\ref{sec:edr}-C, future EDR-blockchain architectures should decouple \emph{response} (which must be fast, and can occur off-chain via conventional orchestration) from \emph{audit and accountability} (which can tolerate consensus latency and is precisely where blockchain's immutability is valuable). Concretely: an endpoint agent executes containment immediately using local policy, then asynchronously commits a tamper-evident record of the action to the ledger for later audit, dispute resolution, and cross-organization accountability. This mirrors the R1$\rightarrow$R2 transition pattern already visible in the IDS literature (Table~\ref{tab:taxonomy}) but applied specifically to endpoint telemetry rather than network flows.

\subsection{Lightweight, EDR-Appropriate Consensus}

The consensus-protocol innovation already underway for IDS settings (e.g., PoFQ~\cite{hieu2025pofq}) should be extended and benchmarked specifically against EDR-realistic latency budgets (sub-second to low-single-digit-second containment windows), rather than the more latency-tolerant budgets typical of periodic FL-aggregation rounds. This requires a dedicated EDR-specific benchmark suite, which does not currently exist in the surveyed literature.

\subsection{Post-Quantum-Ready Detection-and-Response Ledgers}

Given the cryptographic exposure identified in Section~\ref{sec:crosscutting}-B, and the long deployment lifecycles typical of industrial and vehicular detection infrastructure, future blockchain-IDS/EDR architectures should incorporate cryptographic agility (the ability to swap signature/hash primitives without redesigning the ledger) from the outset, drawing directly on the post-quantum blockchain literature~\cite{yang2023survey,parida2023post,baseri2025blockchain} rather than treating this as a future migration problem.

\subsection{Smart-Contract Security as a First-Class Design Concern for Response Automation}

Any R2/R3 system that uses a smart contract to trigger containment should undergo the same threat-modeling rigor applied to smart contracts in financial (DeFi) contexts~\cite{liu2025blockchain}, including reentrancy-style analysis adapted to the security-response domain (e.g., could a compromised endpoint trigger a false containment action against a legitimate node to cause denial-of-service?).

\subsection{Standardized, Latency-Aware Evaluation Benchmarks}

The field would benefit from a shared benchmark that reports detection accuracy \emph{jointly} with end-to-end response latency under realistic device/network constraints, rather than as separately reported, decoupled metrics. This would make the accuracy-overhead tension noted in Section~\ref{sec:crosscutting}-D visible and comparable across papers rather than implicit within each.

\subsection{Cross-Chain Interoperability for Multi-Stakeholder Detection Consortia}

Future work should explore interoperability protocols allowing independently governed detection ledgers (e.g., one per healthcare network, one per fleet operator) to share threat intelligence in a verifiable way without requiring a single shared chain, extending cross-chain interoperability research into the IDS/EDR domain specifically.

\subsection{LLM-Based Detection Engines for Blockchain-Anchored EDR}

Given the absence of LLM-based detection engines in any EDR-class system surveyed (Section~\ref{sec:ai-engine}), and given LLMs' demonstrated suitability for synthesizing the heterogeneous, semi-structured telemetry that endpoint agents collect, an LLM-based detection core paired with the hybrid on-chain/off-chain response architecture proposed above (Section~\ref{sec:future}.A) is a natural and currently unexplored research direction. Designs in this space should explicitly evaluate adversarial robustness against prompt-injection-style manipulation of the detection model (Section~\ref{sec:crosscutting}.C), since an EDR system's response actions carry direct operational consequences (host isolation, process termination) that make a successfully manipulated verdict considerably more costly than in a purely advisory detection setting.

\subsection{Extending Beyond EDR Toward XDR}

Given that even EDR-blockchain integration is nascent, XDR (which additionally correlates identity, cloud, and network telemetry) represents a longer-horizon direction. The functional-role taxonomy developed here (Section~\ref{sec:taxonomy}) should transfer directly, but the trust-and-consensus questions become more complex once multiple distinct telemetry domains, potentially governed by different organizational stakeholders, must be reconciled on a shared or interoperating ledger.

\section{Conclusion}\label{sec:conclusion}
This paper presents a Systematization of Knowledge (SoK) that fundamentally reframes the intersection of blockchain technology and intrusion detection paradigms. By moving beyond traditional chronological or domain-specific cataloging, we introduce a three-axis taxonomy that explicitly decouples blockchain's functional architectural roles\textemdash namely evidentiary logging, decentralized consensus, incentive alignment, and federated learning integrity assurance\textemdash from the underlying detection classes and response maturities. Our systematic review of the literature from 2018 through 2026 uncovers a stark architectural asymmetry: while blockchain-backed network-level detection (NIDS) is highly saturated, its integration into Endpoint Detection and Response (EDR) and Extended Detection and Response (XDR) pipelines remains substantially unexamined. 

We have analyzed the structural root causes of this gap, highlighting critical mismatches in operational latency, vendor deployment models, and disciplinary community isolation. Furthermore, this work exposes systemic cross-cutting vulnerabilities that the current literature treats as peripheral, notably the immediate threat of quantum cryptanalysis to legacy ledgers, smart-contract execution attack surfaces, and the emerging adversarial risks of Large Language Model (LLM)-based detection engines. To advance past these limitations, we lay out a clear research roadmap prioritizing hybrid on-chain/off-chain response architectures and post-quantum agile cryptographic layers. Ultimately, this systematization serves as both a diagnostic assessment of the field's current imbalances and an architectural blueprint for engineering next-generation, decentralized, and resilient cyber defense ecosystems.

\bibliographystyle{IEEEtran}
\bibliography{Ref.bib} 

@article{hassan2023survey,
  title={A survey on blockchain-based intrusion detection systems for iot},
  author={Hassan, Jawad and Abid, Muhammad Kamran and Ghulam, Ali and Fakhar, Muhammad Salman and Asif, Muhammad and others},
  journal={VAWKUM Transactions on Computer Sciences},
  volume={11},
  number={1},
  pages={138--151},
  year={2023}
}

@article{shalabi2024blockchain,
  title={A blockchain-based intrusion detection/prevention systems in IoT network: a systematic review},
  author={Shalabi, Khawla and Al-Haija, Qasem Abu and Al-Fayoumi, Mustafa},
  journal={Procedia Computer Science},
  volume={236},
  pages={410--419},
  year={2024},
  publisher={Elsevier}
}

@article{asante2025blockchain,
  title={Blockchain and deep learning for improving impenetrable data passage in the IoT-imbued healthcare system},
  author={Asante, Isaac Osei and Wu, Libing},
  journal={Cluster Computing},
  volume={28},
  number={7},
  pages={481},
  year={2025},
  publisher={Springer}
}

@inproceedings{li2020framework,
  title={A framework of blockchain-based collaborative intrusion detection in software defined networking},
  author={Li, Wenjuan and Tan, Jiao and Wang, Yu},
  booktitle={International Conference on Network and System Security},
  pages={261--276},
  year={2020},
  organization={Springer}
}

@article{li2023blockchain,
  title={A blockchain-enabled collaborative intrusion detection framework for SDN-assisted cyber-physical systems},
  author={Li, Wenjuan and Wang, Yu and Li, Jin},
  journal={International Journal of Information Security},
  volume={22},
  number={5},
  pages={1219--1230},
  year={2023},
  publisher={Springer Nature BV}
}

@article{almuseelem2026collaborative,
  title={Collaborative intrusion detection systems in SD-WAN enterprise networks with federated learning: a comprehensive survey},
  author={Almuseelem, Waleed},
  journal={Journal of King Saud University Computer and Information Sciences},
  year={2026},
  publisher={Springer}
}

@article{nishad2026lightweight,
  title={A lightweight blockchain-enabled federated intrusion prevention framework for resource-constrained industrial IoT devices to detect and mitigate emerging cyberattacks},
  author={Nishad, Dinesh Kumar and Singh, Rashmi and Khalid, Saifullah},
  journal={Journal of Cloud Computing},
  year={2026},
  publisher={Springer}
}

@article{kumar2022p2tif,
  title={P2tif: A blockchain and deep learning framework for privacy-preserved threat intelligence in industrial iot},
  author={Kumar, Prabhat and Kumar, Randhir and Gupta, Govind P and Tripathi, Rakesh and Srivastava, Gautam},
  journal={IEEE transactions on industrial informatics},
  volume={18},
  number={9},
  pages={6358--6367},
  year={2022},
  publisher={IEEE}
}

@article{liu2021artificial,
  title={Artificial intelligence-enabled DDoS detection for blockchain-based smart transport systems},
  author={Liu, Tong and Sabrina, Fariza and Jang-Jaccard, Julian and Xu, Wen and Wei, Yuanyuan},
  journal={Sensors},
  volume={22},
  number={1},
  pages={32},
  year={2021},
  publisher={MDPI}
}

@article{khalil2022blockchain,
  title={A blockchain footprint for authentication of IoT-enabled smart devices in smart cities: State-of-the-art advancements, challenges and future research directions},
  author={Khalil, Usman and Malik, Owais Ahmed and Hussain, Saddam and others},
  journal={IEEE Access},
  volume={10},
  pages={76805--76823},
  year={2022},
  publisher={IEEE}
}

@article{meng2018intrusion,
  title={When intrusion detection meets blockchain technology: a review},
  author={Meng, Weizhi and Tischhauser, Elmar Wolfgang and Wang, Qingju and Wang, Yu and Han, Jinguang},
  journal={Ieee Access},
  volume={6},
  pages={10179--10188},
  year={2018},
  publisher={IEEE}
}

@article{mathew2022integration,
  title={Integration of blockchain and collaborative intrusion detection for secure data transactions in industrial IoT: a survey},
  author={Mathew, Sujith Samuel and Hayawi, Kadhim and Dawit, Nahom Aron and Taleb, Ikbal and Trabelsi, Zouheir},
  journal={Cluster Computing},
  volume={25},
  number={6},
  pages={4129--4149},
  year={2022},
  publisher={Springer}
}

@article{peelam2025blockchain,
  title={Blockchain-enabled intrusion detection systems for real-time vehicle monitoring},
  author={Peelam, Mritunjay Shall and Chamola, Vinay and Chaurasia, Brijesh Kumar},
  journal={Vehicular Communications},
  pages={100961},
  year={2025},
  publisher={Elsevier}
}

@article{abdel2021federated,
  title={Federated intrusion detection in blockchain-based smart transportation systems},
  author={Abdel-Basset, Mohamed and Moustafa, Nour and Hawash, Hossam and Razzak, Imran and Sallam, Karam M and Elkomy, Osama M},
  journal={IEEE Transactions on Intelligent Transportation Systems},
  volume={23},
  number={3},
  pages={2523--2537},
  year={2021},
  publisher={IEEE}
}

@inproceedings{kaur2024evolution,
  title={Evolution of endpoint detection and response (edr) in cyber security: A comprehensive review},
  author={Kaur, Harpreet and Sanjaiy SL, Dharani and Paul, Tirtharaj and Kumar Thakur, Rohit and Kumar Reddy, K Vijay and Mahato, Jay and Naveen, Kaviti},
  booktitle={E3S Web of Conferences},
  volume={556},
  pages={01006},
  year={2024},
  organization={EDP Sciences}
}

@inproceedings{eigner2026comparison,
  title={Comparison of the Endpoint Detection and Response (EDR) Solutions From CrowdStrike and SentinelOne},
  author={Eigner, Christoph and Fahrnberger, G{\"u}nter},
  booktitle={International Conference on Innovations for Community Services},
  pages={157--182},
  year={2026},
  organization={Springer}
}

@article{aliyu2025blockchain,
  title={A Blockchain-Enhanced Deep Learning Approach for Intrusion Detection in Trusted Execution Environments},
  author={Aliyu, Ahmed Abubakar and Ibrahim, Mohammed and Abdulkadir, Sa'adatu},
  journal={Digital Technologies Research and Applications},
  volume={4},
  number={1},
  pages={135--157},
  year={2025}
}

@article{li2021surveying,
  title={Surveying trust-based collaborative intrusion detection: state-of-the-art, challenges and future directions},
  author={Li, Wenjuan and Meng, Weizhi and Kwok, Lam For},
  journal={IEEE Communications Surveys \& Tutorials},
  volume={24},
  number={1},
  pages={280--305},
  year={2021},
  publisher={IEEE}
}

@inproceedings{fan2022blockchain,
  title={A blockchain-based retribution mechanism for collaborative intrusion detection},
  author={Fan, Wenjun and Kumar, Shubham and Chang, Sang-Yoon and Park, Younghee},
  booktitle={Silicon Valley Cybersecurity Conference},
  pages={57--73},
  year={2022},
  organization={Springer}
}

@article{begum2024bflids,
  title={BFLIDS: Blockchain-driven federated learning for intrusion detection in IoMT networks},
  author={Begum, Khadija and Mozumder, Md Ariful Islam and Joo, Moon-Il and Kim, Hee-Cheol},
  journal={Sensors},
  volume={24},
  number={14},
  pages={4591},
  year={2024},
  publisher={MDPI}
}

@article{miller2024collaborative,
  title={Collaborative Cybersecurity Using Blockchain: A Survey},
  author={Miller, Lo{\"\i}c and Pahl, Marc-Oliver},
  journal={arXiv preprint arXiv:2403.04410},
  year={2024}
}

@article{alevizos2025automated,
  title={Automated cybersecurity compliance and threat response using AI, blockchain and smart contracts},
  author={Alevizos, Lampis},
  journal={International Journal of Information Technology},
  volume={17},
  number={2},
  pages={767--781},
  year={2025},
  publisher={Springer}
}

@inproceedings{kolokotronis2019blockchain,
  title={On blockchain architectures for trust-based collaborative intrusion detection},
  author={Kolokotronis, Nicholas and Brotsis, Sotirios and Germanos, Georgios and Vassilakis, Costas and Shiaeles, Stavros},
  booktitle={2019 IEEE world congress on services (SERVICES)},
  volume={2642},
  pages={21--28},
  year={2019},
  organization={IEEE}
}

@article{selimi2023cybernfts,
  title={CyberNFTs: conceptualising a decentralised and reward-driven intrusion detection system with ML},
  author={Selimi, Synim and Rexha, Blerim and Vishi, Kamer},
  journal={International Journal of Information and Computer Security},
  volume={22},
  number={1},
  pages={117--138},
  year={2023},
  publisher={Inderscience Publishers (IEL)}
}

@article{ali2025federated,
  title={Federated learning-enhanced blockchain framework for privacy-preserving intrusion detection in industrial iot},
  author={Ali, Anas and Husain, Mubashar and Hans, Peter},
  journal={arXiv preprint arXiv:2505.15376},
  year={2025}
}

@inproceedings{al2022blockchain,
  title={Blockchain-based monitoring for poison attack detection in decentralized federated learning},
  author={Al Mallah, Ranwa and Lopez, David},
  booktitle={2022 International Conference on Electrical, Computer, Communications and Mechatronics Engineering (ICECCME)},
  pages={1--6},
  year={2022},
  organization={IEEE}
}

@article{martinez2021host,
  title={A host intrusion detection system architecture for embedded industrial devices},
  author={Martinez, Cyntia Vargas and Vogel-Heuser, Birgit},
  journal={journal of the Franklin Institute},
  volume={358},
  number={1},
  pages={210--236},
  year={2021},
  publisher={Elsevier}
}

@article{subba2021tfidfvectorizer,
  title={A tfidfvectorizer and singular value decomposition based host intrusion detection system framework for detecting anomalous system processes},
  author={Subba, Basant and Gupta, Prakriti},
  journal={Computers \& Security},
  volume={100},
  pages={102084},
  year={2021},
  publisher={Elsevier}
}

@inproceedings{rahul2025blockchain,
  title={Blockchain Integrated Intelligent Firewall System for Real Time Intrusion Detection},
  author={Rahul, R and Sandhya, V and Yasvanth, A and Varun, E and others},
  booktitle={2025 4th International Conference on Advances in Computing, Communication, Embedded and Secure Systems (ACCESS)},
  pages={350--356},
  year={2025},
  organization={IEEE}
}

@inproceedings{chan2023complete,
  title={A Complete Log Files Security Solution Using Anomaly Detection and Blockchain Technology},
  author={Chan, Tshun Kong and Kamsin, Intan Farahama Binti and Amin, Salmiah and Zainal, Nur Khairunnisha},
  booktitle={2023 15th International Conference on Developments in eSystems Engineering (DeSE)},
  pages={112--117},
  year={2023},
  organization={IEEE}
}

@article{hieu2025pofq,
  title={PoFQ: a blockchain consensus protocol for decentralized federated learning-based threat hunting approach in a trustless computing landscape},
  author={Hieu, Vo Truong Trung and Quyen, Nguyen Huu and Do Hoang, Hien and Duy, Phan The and Pham, Van-Hau},
  journal={Cluster Computing},
  volume={28},
  number={9},
  pages={571},
  year={2025},
  publisher={Springer}
}

@article{ar2025enhancing,
  title={Enhancing anomaly detection and prevention in Internet of Things (IoT) using deep neural networks and blockchain based cyber security},
  author={AR, Sathyabama and Katiravan, Jeevaa},
  journal={Scientific Reports},
  volume={15},
  number={1},
  pages={22369},
  year={2025},
  publisher={Nature Publishing Group UK London}
}

@article{liu2025blockchain,
  title={Blockchain smart contract security: Threats and mitigation strategies in a lifecycle perspective},
  author={Liu, Detian and Zhang, Jianbiao and Wang, Yifan and Shen, Hong and Zhang, Zhaoqian and Ye, Tao},
  journal={ACM Computing Surveys},
  volume={58},
  number={4},
  pages={1--34},
  year={2025},
  publisher={ACM New York, NY}
}

@inproceedings{adele2024comprehensive,
  title={A comprehensive systematic review of blockchain-based intrusion detection systems},
  author={Adele, Gideon and Borah, Abinash and Paranjothi, Anirudh and Khan, Mohammad S and Poulkov, Vladimir K},
  booktitle={2024 IEEE World AI IoT Congress (AIIoT)},
  pages={605--611},
  year={2024},
  organization={IEEE}
}

@article{liu2021blockchain,
  title={Blockchain and federated learning for collaborative intrusion detection in vehicular edge computing},
  author={Liu, Hong and Zhang, Shuaipeng and Zhang, Pengfei and Zhou, Xinqiang and Shao, Xuebin and Pu, Geguang and Zhang, Yan},
  journal={IEEE Transactions on Vehicular Technology},
  volume={70},
  number={6},
  pages={6073--6084},
  year={2021},
  publisher={IEEE}
}

@article{he2022cgan,
  title={CGAN-based collaborative intrusion detection for UAV networks: A blockchain-empowered distributed federated learning approach},
  author={He, Xiaoqiang and Chen, Qianbin and Tang, Lun and Wang, Weili and Liu, Tong},
  journal={IEEE Internet of Things Journal},
  volume={10},
  number={1},
  pages={120--132},
  year={2022},
  publisher={IEEE}
}

@inproceedings{abou2023securing,
  title={Securing federated learning through blockchain and explainable AI for robust intrusion detection in IoT networks},
  author={Abou El Houda, Zakaria and Moudoud, Hajar and Brik, Bouziane and Khoukhi, Lyes},
  booktitle={IEEE INFOCOM 2023-IEEE Conference on Computer Communications Workshops (INFOCOM WKSHPS)},
  pages={1--6},
  year={2023},
  organization={IEEE}
}

@article{huang2025adaptive,
  title={An adaptive intrusion detection system for the internet of things using large language models and post-quantum-secure blockchain},
  author={Huang, Yunfan and Ma, Maode and Raymond, Wong Jee Keen and Chow, Chee-Onn},
  journal={Computer Networks},
  pages={111819},
  year={2025},
  publisher={Elsevier}
}

@article{ayadi2026blockchain,
  title={A Blockchain-Enabled Smart Contract Architecture for Enhancing Transparency, Traceability, and Trust in Global Supply Chain Management},
  author={Ayadi, Naim and Hussain, Syed Arshad and Deen, Arif and Ullah, Asadullah and Hakro, Dil Nawaz and Babar, Muhammad and Jariko, Mushtaque Ali and Al Farsi, Alya and Hussain, Akhtar},
  journal={Computers},
  volume={15},
  number={3},
  pages={198},
  year={2026},
  publisher={MDPI}
}

@article{sun2024blockchain,
  title={Blockchain based federated learning for intrusion detection for Internet of Things},
  author={Sun, Nan and Wang, Wei and Tong, Yongxin and Liu, Kexin},
  journal={Frontiers of Computer Science},
  volume={18},
  number={5},
  pages={185328},
  year={2024},
  publisher={Springer}
}

@article{cao2025privacy,
  title={Privacy and Trust in Blockchain-Federated Intrusion Detection Systems: Taxonomy, Challenges and Perspectives},
  author={Cao, Yuan and Ku, Chin Soon and Kumar, Rahul and Khan, Arshad},
  journal={Journal of Reliable and Secure Computing},
  volume={1},
  number={1},
  pages={4--24},
  year={2025},
  publisher={Institute of Central Computation and Knowledge}
}

@inproceedings{zaabar2022intrusion,
  title={Intrusion detection system for IoMT through blockchain-based federated learning},
  author={Zaabar, Bessem and Cheikhrouhou, Omar and Abid, Mohamed},
  booktitle={2022 15th International Conference on Security of Information and Networks (SIN)},
  pages={01--08},
  year={2022},
  organization={IEEE}
}

@article{shalan2025enhancing,
  title={Enhancing smart home security: blockchain-enabled federated learning with knowledge distillation for intrusion detection},
  author={Shalan, Mohammed and Hasan, Md Rakibul and Bai, Yan and Li, Juan},
  journal={Smart Cities},
  volume={8},
  number={1},
  pages={35},
  year={2025},
  publisher={MDPI}
}

@article{qu2022blockchain,
  title={Blockchain-enabled federated learning: A survey},
  author={Qu, Youyang and Uddin, Md Palash and Gan, Chenquan and Xiang, Yong and Gao, Longxiang and Yearwood, John},
  journal={ACM Computing Surveys},
  volume={55},
  number={4},
  pages={1--35},
  year={2022},
  publisher={ACM New York, NY}
}

@article{wu2023survey,
  title={A survey on blockchain-based federated learning},
  author={Wu, Lang and Ruan, Weijian and Hu, Jinhui and He, Yaobin},
  journal={Future Internet},
  volume={15},
  number={12},
  pages={400},
  year={2023},
  publisher={MDPI}
}

@article{nguyen2021federated,
  title={Federated learning meets blockchain in edge computing: Opportunities and challenges},
  author={Nguyen, Dinh C and Ding, Ming and Pham, Quoc-Viet and Pathirana, Pubudu N and Le, Long Bao and Seneviratne, Aruna and Li, Jun and Niyato, Dusit and Poor, H Vincent},
  journal={IEEE Internet of Things Journal},
  volume={8},
  number={16},
  pages={12806--12825},
  year={2021},
  publisher={IEEE}
}

@article{li2021blockchain,
  title={Blockchain assisted decentralized federated learning (BLADE-FL): Performance analysis and resource allocation},
  author={Li, Jun and Shao, Yumeng and Wei, Kang and Ding, Ming and Ma, Chuan and Shi, Long and Han, Zhu and Poor, H Vincent},
  journal={IEEE Transactions on Parallel and Distributed Systems},
  volume={33},
  number={10},
  pages={2401--2415},
  year={2021},
  publisher={IEEE}
}

@article{lu2020blockchain,
  title={Blockchain empowered asynchronous federated learning for secure data sharing in internet of vehicles},
  author={Lu, Yunlong and Huang, Xiaohong and Zhang, Ke and Maharjan, Sabita and Zhang, Yan},
  journal={IEEE Transactions on Vehicular Technology},
  volume={69},
  number={4},
  pages={4298--4311},
  year={2020},
  publisher={IEEE}
}

@article{elumalai2026blockchain,
  title={A blockchain-driven intrusion detection model for secure communication in IoT-WSN mesh architectures},
  author={Elumalai, G and Kumar, J Arun and Sivakumar, P and Teresa, VV},
  journal={Peer-to-Peer Networking and Applications},
  volume={19},
  number={2},
  pages={60},
  year={2026},
  publisher={Springer}
}

@article{ilyas2023prevention,
  title={Prevention of DDoS attacks using an optimized deep learning approach in blockchain technology},
  author={Ilyas, Benkhaddra and Kumar, Abhishek and Setitra, Mohamed Ali and Bensalem, ZineEl Abidine and Lei, Hang},
  journal={Transactions on Emerging Telecommunications Technologies},
  volume={34},
  number={4},
  pages={e4729},
  year={2023},
  publisher={Wiley Online Library}
}

@article{jiang2024credible,
  title={Credible link flooding attack detection and mitigation: A blockchain-based approach},
  author={Jiang, Xiaofeng and Shi, Qianbao and Miao, Hengkun and Cao, Wanqin and He, Huasen and Chen, Shuangwu and Yang, Jian},
  journal={IEEE Transactions on Network and Service Management},
  volume={21},
  number={3},
  pages={3537--3554},
  year={2024},
  publisher={IEEE}
}

@article{zhang2025attacks,
  title={DoS Attacks and Defense Technologies in Blockchain Systems: A Hierarchical Analysis},
  author={Zhang, Chunyi and Dou, Fengjiao and Li, Xiaoqi},
  journal={arXiv preprint arXiv:2507.22611},
  year={2025}
}

@article{song2025hybrid,
  title={A hybrid blockchain and machine learning approach for intrusion detection system in Industrial Internet of Things},
  author={Song, Wu and Zhu, Xiangyuan and Ren, Sheng and Tan, Wenxue and Peng, Yibo},
  journal={Alexandria Engineering Journal},
  volume={127},
  pages={619--627},
  year={2025},
  publisher={Elsevier}
}

@article{jaffal2025large,
  title={Large language models in cybersecurity: A survey of applications, vulnerabilities, and defense techniques},
  author={Jaffal, Niveen O and Alkhanafseh, Mohammed and Mohaisen, David},
  journal={AI},
  volume={6},
  number={9},
  pages={216},
  year={2025},
  publisher={MDPI}
}

@article{yang2025large,
  title={Large language models for network intrusion detection systems: Foundations, implementations, and future directions},
  author={Yang, Shuo and Zheng, Xinran and Zhang, Xinchen and Xu, Jinfeng and Li, Jinze and Xie, Donglin and Long, Weicai and Ngai, Edith CH},
  journal={arXiv preprint arXiv:2507.04752},
  year={2025}
}

@article{jamshidi2026tri,
  title={Tri-LLM Cooperative Federated Zero-Shot Intrusion Detection with Semantic Disagreement and Trust-Aware Aggregation},
  author={Jamshidi, Saeid and Wahab, Omar Abdul and Khomh, Foutse and Nafi, Kawser Wazed},
  journal={arXiv preprint arXiv:2602.00219},
  year={2026}
}

@article{luo2023bc4llm,
  title={Bc4llm: Trusted artificial intelligence when blockchain meets large language models},
  author={Luo, Haoxiang and Luo, Jian and Vasilakos, Athanasios V},
  journal={arXiv preprint arXiv:2310.06278},
  year={2023}
}

@article{geren2025blockchain,
  title={Blockchain for large language model security and safety: A holistic survey},
  author={Geren, Caleb and Board, Amanda and Dagher, Gaby G and Andersen, Tim and Zhuang, Jun},
  journal={ACM SIGKDD explorations newsletter},
  volume={26},
  number={2},
  pages={1--20},
  year={2025},
  publisher={ACM New York, NY, USA}
}

@article{orabi2025adapting,
  title={Adapting security and decentralized knowledge enhancement in federated learning using blockchain technology: literature review},
  author={Orabi, Menna Mamdouh and Emam, Osama and Fahmy, Hanan},
  journal={Journal of Big Data},
  volume={12},
  number={1},
  pages={55},
  year={2025},
  publisher={Springer}
}

@article{parmar2026zer0n,
  title={Zer0n: An AI-Assisted Vulnerability Discovery and Blockchain-Backed Integrity Framework},
  author={Parmar, Harshil and Vyas, Pushti and Khristi, Prayers and Panchal, Priyank},
  journal={arXiv preprint arXiv:2601.07019},
  year={2026}
}

@article{ieropoulos2026empowering,
  title={Empowering IoT Security: On-Device Intrusion Detection in Resource Constrained Devices},
  author={Ieropoulos, Vasilis and Anthi, Eirini and Spyridopoulos, Theodoros and Burnap, Pete and Khan, Aftab and Carnelli, Pietro},
  journal={arXiv preprint arXiv:2605.13159},
  year={2026}
}

@article{gharavi2024post,
  title={Post-quantum blockchain security for the Internet of Things: Survey and research directions},
  author={Gharavi, Hadi and Granjal, Jorge and Monteiro, Edmundo},
  journal={IEEE Communications Surveys \& Tutorials},
  volume={26},
  number={3},
  pages={1748--1774},
  year={2024},
  publisher={IEEE}
}

@article{yang2023survey,
  title={A survey and comparison of post-quantum and quantum blockchains},
  author={Yang, Zebo and Alfauri, Haneen and Farkiani, Behrooz and Jain, Raj and Di Pietro, Roberto and Erbad, Aiman},
  journal={IEEE Communications Surveys \& Tutorials},
  volume={26},
  number={2},
  pages={967--1002},
  year={2023},
  publisher={IEEE}
}

@article{baseri2025blockchain,
  title={Blockchain security risk assessment in quantum era, migration strategies and proactive defense},
  author={Baseri, Yaser and Hafid, Abdelhakim and Shahsavari, Yahya and Makrakis, Dimitrios and Khodaiemehr, Hassan},
  journal={IEEE Communications Surveys \& Tutorials},
  year={2025},
  publisher={IEEE}
}

@article{khan2025securing,
  title={Securing IOT in the post-quantum era: implementation, challenges and future directions},
  author={Khan, Muhammad Asghar and Khan, Muhammad Attique},
  journal={IEEE Communications Standards Magazine},
  year={2025},
  publisher={IEEE}
}

@article{parida2023post,
  title={Post-quantum distributed ledger technology: a systematic survey},
  author={Parida, Nikhil Kumar and Jatoth, Chandrashekar and Reddy, V Dinesh and Hussain, Md Muzakkir and Faizi, Jamilurahman},
  journal={Scientific Reports},
  volume={13},
  number={1},
  pages={20729},
  year={2023},
  publisher={Nature Publishing Group UK London}
}

@article{chidambaranathan2024deep,
  title={Deep learning enabled blockchain based electronic heathcare data attack detection for smart health systems},
  author={Chidambaranathan, S and Geetha, R},
  journal={Measurement: Sensors},
  volume={31},
  pages={100959},
  year={2024},
  publisher={Elsevier}
}

@article{shahsavari2026sok,
  title={SoK: Federated Learning for Intrusion Detection in Vehicular Networks},
  author={Shahsavari, Yahya and Nourmohammadi, Reza and Rouhani, Sara and Zhang, Kaiwen},
  journal={arXiv preprint arXiv:2607.10914},
  year={2026}
}
\end{document}